\documentclass{elsarticle}
\usepackage{amsmath}
\usepackage{amssymb}
\usepackage{accents}

\usepackage{graphicx}
\usepackage{dcolumn}
\usepackage{bm}
\usepackage{cancel}
\usepackage{color}
\usepackage[utf8]{inputenc}

\begin{document}

\newcommand{\nbar}{\underaccent{\bar}{n}}
\newcommand{\mubar}{\underaccent{\bar}{\mu}}
\newcommand{\cbar}{\underaccent{\bar}{c}}

\title{General statistics of particles without spin}

\author{M. Hoyuelos}
\ead{hoyuelos@mdp.edu.ar}
\address{Instituto de Investigaciones F\'isicas de Mar del Plata (IFIMAR-CONICET) and Departamento de F\'isica, Facultad de Ciencias Exactas y Naturales, Universidad Nacional de Mar del Plata, Funes 3350, 7600 Mar del Plata, Argentina}

\begin{abstract}
The possibility of obtaining exotic statistics, different from Bose-Einstein or Fermi-Dirac, is analyzed, in the context of quantum field theory, through the inclusion of a counting operator in the definition of the partition function. This operator represents the statistical weight of the allowed number states. In particular, the statistics of ewkons (introduced in Phys. Rev. E \textbf{94}, 062115, 2016) is analyzed. They have a negative relation between pressure and energy density, a feature shared by dark energy. The present approach develops the possibility of describing dark energy with a field of non-interacting particles with ewkon's statistics. 

\end{abstract}

\begin{keyword}
quantum statistics \sep partition function \sep dark energy \sep exotic statistics
\end{keyword}


\maketitle

\section{Introduction}

Quantum statistics for bosons and fermions are derived in the context of finite temperature field theory from the corresponding free Lagrangian; see, for example, \cite{kapusta,laine}. 
The main point that I wish to address in this paper is whether other kind of statistics, introduced through different statistical weights, may be useful to describe nature. The motivation of this search lies in dark energy. The observed accelerated expansion of the universe implies a negative pressure that is attributed to dark energy \cite{hogan,planck}. But an ideal gas of fermions or bosons has positive pressure; it is necessary to assume some kind of unknown interaction in the Lagrangian to solve this problem \cite{amendola,gott}. Here I consider a different approach. We can keep the free Lagrangian of non-interacting particles but take into account the possibility of statistics other than the corresponding to fermions or bosons. The possibility that a negative pressure can be obtained from non-interacting particles with an appropriate statistics was recently analyzed in reference \cite{hoyusist}. The conjecture that non-interacting particles have free diffusion in energy space was shown to lead to the known distributions of Fermi-Dirac, Bose-Einstein and Maxwell-Boltzmann, and to a new one: ewkons. Occupation number for ewkons has the shape of an exponential, like for classical particles, but shifted a given positive quantity. They have negative pressure \cite{hoyusist,hoyuelos}; moreover, the relation between pressure and energy density is close to $-1$ [see Eq.\ (40) in Ref.\ \cite{hoyuelos} for a non-relativistic gas of ewkons]. There are several examples of previous works on exotic or intermediate statistics, that go beyond fermions or bosons, useful to describe specific systems, in some cases as a way to incorporate interactions; see, for example \cite{gentile,green,katsura,wilczek,greenberg,haldane,isakov,isakov2,poly,anghel,dai}; for a review, see \cite{khare}.

I consider a system composed by non-interacting particles without spin, described by a scalar field $\phi$ that obeys the Klein-Gordon equation. The assumption of equal statistical weights for the allowed number states implies that such system in equilibrium has occupation number with the Bose-Einstein distribution. Nevertheless, there may be situations in which that assumption is not appropriate. In Sect.\ \ref{variable}, I analyze the different statistical weights of number states that lead to the aforementioned distributions for bosons, fermions, classical particles and ewkons.

It is possible to reproduce such distributions evaluating the partition function in the context of quantum field theory, that is, evaluating the trace in the base of eigenstates of the field operator. A first step is accomplished in Sect.\ \ref{0+1} where the partition function for the harmonic oscillator is obtained; this is equivalent to a $0+1$ dimension theory. The extension to a $d+1$ dimensional spacetime is presented in Sect.\ \ref{d+1}. Conclusions are worked out in Sect.\ \ref{conclusions}.

\section{Counting operator}
\label{variable}

Let us consider a system in contact with a heat reservoir at temperature $T$. Besides heat, system and reservoir can also exchange particles. In the grand canonical ensemble it is assumed that the reservoir is also a reservoir of particles that imposes a constant chemical potential per particle $\mu$. For a quantum system with Hamiltonian $\hat{H}$ and number of particles operator $\hat{n}$, the grand partition function is
\begin{equation}
\mathcal{Z} = \text{tr}\, e^{-\beta (\hat{H}- \mu \hat{n})}
\label{defZ}
\end{equation}
where the trace is evaluated using a base of normalized states. Let us consider non-interacting particles in a harmonic oscillator with frequency $\omega$ as a preliminary step before dealing with a quantum field, that can be thought as a set of infinite harmonic oscillators. The Hamiltonian is $\hat{H} = (\hat{n}+1/2)\hbar \omega$. The grand partition function is obtained evaluating the trace in the base of number eigenstates:
\begin{equation}
\mathcal{Z} = e^{-\beta \hbar \omega/2} \sum_{n=0}^{\infty} \delta_n e^{-\beta(\hbar \omega-\mu)n}.
\label{partfun}
\end{equation}
where $\delta_n$ is a counting factor included to represent, in principle, both statistics of Bose-Einstein and Fermi-Dirac in the same equation. For the first case, $\delta_n = 1$ $\forall n$; and for the second, $\delta_n = 1$ for $n=0$ or 1, and $\delta_n=0$ for $n\ge 2$. The definition \eqref{defZ} remains unchanged for these two canonical cases when the counting operator $\hat{\delta}$, introduced in Ref.\ \cite{hoyuelos}, is included:
\begin{equation}
\mathcal{Z} = \text{tr}\,[ \hat{\delta} e^{-\beta (\hat{H}- \mu \hat{n})}]
\label{qpartfun}
\end{equation}
Operator $\hat{\delta}$ commutes with $\hat{n}$ and has eigenvalues equal to the counting factor mentioned before: $\langle n |\hat{\delta} |n\rangle = \delta_n$. Here I wish to explore the possibility of other kinds of particles that require eigenvalues $\delta_n$ different from 1 or 0 for their statistical description. For example, Eq.\ \eqref{qpartfun} can also represent Maxwell-Boltzmann statistics, also called Quantum Boltzmann statistics in a quantum mechanical context \cite{isakov}; in this case we have $\delta_n = 1/n!$, this is equivalent to consider a not normalized set of number states $|\tilde{n}\rangle = |n\rangle/\sqrt{n!}$ in the definition \eqref{defZ}. The purpose of $\delta_n$, therefore, is not only to determine which states have to be considered in the evaluation of the partition function, but to specify the statistical weight of number states.

I am interested in the statistics of non-interacting particles derived from the condition of free diffusion in energy space \cite{hoyusist}. They correspond to Bose-Einstein, Fermi-Dirac, classical particles, and ewkons. There is no evidence of particles with classical or ewkon statistics at a quantum level. Nevertheless, there is no fundamental principle that forbids the possibility of particles with statistical weight of number states different from 1 or 0, and it may be useful to develop a quantum description of such particles. One reason, as stated in the introduction, is that thermodynamic properties of ewkons share features with those of dark energy.

Ewkons have the Quantum Boltzmann distribution shifted an integer quantity $\sigma$. The ground state is not the vacuum; each energy level has at least $\sigma$ particles. This situation is represented by the following counting factor
\begin{equation}
\delta_n = \left\{ \begin{array}{cl}
0 & \text{if } n<\sigma \\ 
1/(n-\sigma)! & \text{if } n\ge \sigma
\end{array}    \right..
\end{equation}
Then, the grand partition function for ewkons in a harmonic oscillator is
\begin{align}
\mathcal{Z}_\text{ewk} &= e^{-\beta \hbar \omega/2}\sum_{n=\sigma}^\infty \frac{1}{(n-\sigma)!} e^{-\beta(\hbar \omega-\mu)n} \nonumber \\
&= e^{-\beta \hbar \omega/2}\sum_{m=0}^\infty \frac{1}{m!} e^{-\beta(\hbar \omega-\mu)(m+\sigma)} \nonumber  \\
&= e^{-\beta \hbar \omega/2} e^{-\beta(\hbar \omega-\mu)\sigma} \exp\left( e^{-\beta(\hbar \omega-\mu)} \right).
\end{align} 
It is easy to check that, in this case, the mean occupation number is
\begin{equation}
\bar{n} = \frac{1}{\beta} \frac{\partial \ln \mathcal{Z}_\text{ewk}}{\partial \mu} = \sigma + e^{-\beta(\hbar \omega-\mu)}.
\end{equation}

\section{Harmonic oscillator}
\label{0+1}

Before evaluating the partition function in the base of eigenstates of a scalar field operator for a system without interactions, it is useful to do the same in the base of eigenstates of the position operator for the harmonic oscillator. The purpose of this section is to calculate
\begin{equation}
\mathcal{Z} = \int dx\; \langle x| \hat{\delta}\, e^{-\beta \hat{H}} |x\rangle
\label{zxx}
\end{equation}
with $\hat{H} = \frac{1}{2m}\hat{p}^2 + \frac{m\omega^2}{2} \hat{x}^2$; $|x\rangle$ are eigenstates of the position operator $\hat{x}$, $\hat{p}$ is the momentum operator and $m$ is the mass. As usual in quantum field theory, we first consider the case $\mu=0$; the chemical potential can be included later considering both particles and antiparticles; see, e.g., \cite[p. 25]{laine}.  It is possible to calculate the partition function for different statistics in a unified manner, without a priori specifying the eigenvalues of $\hat{\delta}$. First, we write the counting operator in terms of the Hamiltonian:
\begin{align}
\hat{\delta} &= \sum_{n=0}^{\infty} \delta_n |n\rangle \langle n| \nonumber \\
&= \sum_{n=0}^{\infty} \delta_n \int_{0}^{2\pi} \frac{d\varphi}{2\pi} e^{i(\hat{n}-n)\varphi} \nonumber \\
&= \sum_{n=0}^{\infty} \delta_n \int_{0}^{2\pi} \frac{d\varphi}{2\pi} e^{-i(n+1/2)\varphi} e^{i\hat{H}\varphi/\hbar\omega}, 
\label{deltaphi}
\end{align}
where the integral representation of the Kronecker delta was used for $|n\rangle \langle n|$, and $\hat{n}= \hat{H}/\hbar\omega-1/2$. Replacing \eqref{deltaphi} in \eqref{zxx}, we have
\begin{equation}
\mathcal{Z} = \sum_{n=0}^{\infty} \delta_n \int_{0}^{2\pi} \frac{d\varphi}{2\pi} e^{-i(n+1/2)\varphi} \underbrace{\int dx\; \langle x|  e^{-(\beta-i\varphi/\hbar\omega) \hat{H}} |x\rangle}_{\mathcal{Z}'}.
\label{zxx2}
\end{equation}
Let us focus on the integral in $x$:
\begin{equation}
\mathcal{Z}' = \int dx\; \langle x| e^{-(\beta-i\varphi/\hbar\omega) \hat{H}} |x\rangle.
\end{equation}
It can be solved using a standard procedure in which the exponential is spitted in a large number of factors and between each pair another factor  $\int dx_i\;  |x_i\rangle \langle x_i|$ or $\int dp_i\;  |p_i\rangle \langle p_i|$ is inserted, transforming the expression in a path integral. It is well known that, when the term with $\varphi$ in the exponential is absent, this procedure yields the boson's partition function  (see, e.g., \cite[sec.\ 1.1]{laine}):
\begin{equation}
\int dx\; \langle x| e^{-\beta \hat{H}} |x\rangle = \frac{e^{-\beta\hbar\omega/2}}{1-e^{-\beta\hbar\omega}}.
\label{xHx}
\end{equation}
If we repeat the procedure with $\mathcal{Z}'$, the final result is equivalent to make the replacement $\beta \rightarrow \beta-i\varphi/\hbar\omega$ in the previous expression:
\begin{equation}
\mathcal{Z}' = \frac{e^{-(\beta\hbar\omega-i\varphi)/2}}{1-e^{-\beta\hbar\omega+i\varphi}}.
\end{equation}
With this result we go back to Eq.\ \eqref{zxx2} and obtain
\begin{equation}
\mathcal{Z} = e^{-\beta\hbar\omega/2}\sum_{n=0}^{\infty} \delta_n \int_{0}^{2\pi} \frac{d\varphi}{2\pi} \frac{e^{-i n \varphi} }{1-e^{-\beta\hbar\omega+i\varphi}}
\end{equation}
To solve the integral in $\varphi$, we can make the change of variable $z=e^{i\varphi}$ and integrate in the unit circle in the complex plane; the resulting integral can be calculated using the residue theorem. The result 
\begin{equation}
\mathcal{Z} = \sum_{n=0}^{\infty} \delta_n e^{-\beta (n+1/2)\hbar\omega }
\end{equation}
coincides, as expected, with that obtained through the trace with number eigenstates \eqref{partfun} (with $\mu=0$).

\section{Free scalar field}
\label{d+1}

We have now the necessary elements to analyze the possibility of generalized statistics of a free scalar field $\phi$ (eigenvalue of the field operator $\hat{\phi}$). I briefly summarize some basic concepts. In this section I consider units such that $\hbar=1$ and the speed of light is $c=1$. The free Hamiltonian density is 
\begin{equation}
\mathcal{H} = \frac{1}{2} \pi^2 + \frac{1}{2} (\nabla \phi)^2 + \frac{1}{2} m^2 \phi^2
\end{equation}
where $\pi$ is the momentum conjugate of the field, with the canonical commutation relation $[\hat{\phi}(\mathbf{x}),\hat{\pi}(\mathbf{x}')]=i \delta^3(\mathbf{x}-\mathbf{x}')$. It is convenient to use an expansion in Fourier modes:
\begin{equation}
\phi(\mathbf{x}) = \frac{1}{V} \sum_{\mathbf{k}} \phi_\mathbf{k}\, e^{-i\mathbf{k}\cdot\mathbf{x}},
\end{equation}  
where $V$ is the system's volume, and a similar expression for $\pi(\mathbf{x})$. Since $\phi(\mathbf{x}) \in \mathbb{R}$, $\phi_\mathbf{k}^* = \phi_\mathbf{-k}$, and the same for $\pi_\mathbf{k}$. Then, the Hamiltonian is
\begin{align}
H &= \int d\mathbf{x} \; \mathcal{H} \nonumber \\
 &= \frac{1}{V} \sum_{\mathbf{k}} \frac{1}{2}\left( |\pi_\mathbf{k}|^2 + (m^2 + k^2) |\phi_\mathbf{k}|^2 \right)
\end{align}
The system consists in a collection of harmonic oscillators. The Hamiltonian operator can be written as $\hat{H} = \sum_{\mathbf{k}} \hat{H}_\mathbf{k}$ with
\begin{equation}
\hat{H}_\mathbf{k} = (\hat{n}_\mathbf{k} + 1/2) E_\mathbf{k}
\end{equation}
where $E_\mathbf{k} = \sqrt{m^2 + k^2}$, and operator $\hat{n}_\mathbf{k}$ represents the number of quanta in mode $\mathbf{k}$; see the appendix for additional details. For simplicity, I am considering a real scalar field in which only particles have to be taken into account; for a complex scalar field, operators for the number of particles and antiparticles have to be considered.  We define a counting operator $\hat{\delta}_\mathbf{k}$ that represents the statistical weight of a number state $|n_\mathbf{k}\rangle$ in mode $\mathbf{k}$, as we did in the previous section for the harmonic oscillator:
\begin{align}
\hat{\delta}_\mathbf{k} &= \sum_{n_\mathbf{k}=0}^{\infty} \delta_{n_\mathbf{k}} |n_\mathbf{k}\rangle \langle n_\mathbf{k}| \nonumber \\
&= \sum_{n_\mathbf{k}=0}^{\infty} \delta_{n_\mathbf{k}} \int_{0}^{2\pi} \frac{d\varphi_\mathbf{k}}{2\pi} e^{-i(n_\mathbf{k}+1/2)\varphi_\mathbf{k}} e^{i\hat{H}_\mathbf{k}\varphi_\mathbf{k}/E_\mathbf{k}}, 
\label{deltaphik}
\end{align}

The goal is to evaluate the partition function in the base of the field operator eigenstates of a scalar field: 
\begin{equation}
\mathcal{Z}_\text{sf} = \int d\phi \; \langle \phi | \hat{\delta}\, e^{-\beta \hat{H}} |\phi\rangle
\end{equation}
where $\hat{\delta} = \prod_{\mathbf{k}} \hat{\delta}_\mathbf{k}$ represents the statistical weight of the system's number state of $N$ modes that can be written in product form as $|n_{\mathbf{k}_1}\rangle \cdots |n_{\mathbf{k}_N}\rangle$, such that $\hat{\delta} |n_{\mathbf{k}_1}\rangle \cdots |n_{\mathbf{k}_N}\rangle = (\delta_{n_{\mathbf{k}_1}}\cdots \delta_{n_{\mathbf{k}_N}}) |n_{\mathbf{k}_1}\rangle \cdots |n_{\mathbf{k}_N}\rangle$. The field operator eigenstate is also written in product form as $|\phi\rangle = \prod_{\mathbf{k}} |\phi_\mathbf{k}\rangle$, and $d\phi = \prod_{\mathbf{k}} d\phi_\mathbf{k}$. Since different modes are independent, $\hat{\delta}_\mathbf{k}$ acts only on $|\phi_\mathbf{k}\rangle$, and the same for $\hat{H}_\mathbf{k}$. Then, the partition function is
\begin{align}
\mathcal{Z}_\text{sf} &= \prod_{\mathbf{k}} \int d\phi_\mathbf{k}\; \langle \phi_\mathbf{k} | \hat{\delta}_\mathbf{k}\, e^{-\beta \hat{H}_\mathbf{k}} |\phi_\mathbf{k}\rangle \nonumber \\
&= \prod_{\mathbf{k}}\left[ \sum_{n_\mathbf{k}=0}^{\infty} \delta_{n_\mathbf{k}} \int_{0}^{2\pi} \frac{d\varphi_\mathbf{k}}{2\pi} e^{-i(n_\mathbf{k}+1/2)\varphi_\mathbf{k}}
\underbrace{\int d\phi_\mathbf{k}\; \langle \phi_\mathbf{k} |  e^{-(\beta -i\varphi_\mathbf{k}/E_\mathbf{k}) \hat{H}_\mathbf{k}} |\phi_\mathbf{k}\rangle}_{\mathcal{Z}'_\text{sf}} \right],
\end{align} 
where, in the last line, Eq.\ \eqref{deltaphik} was used for $\hat{\delta}_\mathbf{k}$. Now, the procedure goes on as in the case of the harmonic oscillator. The integral in $\phi_\mathbf{k}$, that we call $\mathcal{Z}'_\text{sf}$, can be evaluated using a path integral method. The result is equivalent to make the replacements $\beta \rightarrow \beta - i \varphi_\mathbf{k}/E_\mathbf{k}$ and $\hbar \omega \rightarrow E_\mathbf{k}$ in Eq.\ \eqref{xHx}:
\begin{equation}
\mathcal{Z}'_\text{sf} = \frac{e^{-(\beta E_\mathbf{k} - i\varphi_\mathbf{k})/2}}{1 - e^{-\beta E_\mathbf{k}} e^{i\varphi_\mathbf{k}}}.
\end{equation}
Then, the partition function is
\begin{align}
\mathcal{Z}_\text{sf} &= \prod_{\mathbf{k}} \left[ e^{-\beta E_\mathbf{k}/2} \sum_{n_\mathbf{k}=0}^{\infty} \delta_{n_\mathbf{k}} \int_{0}^{2\pi} \frac{d\varphi_\mathbf{k}}{2\pi} \frac{e^{-i n_\mathbf{k} \varphi_\mathbf{k}}}{1 - e^{-\beta E_\mathbf{k}} e^{i\varphi_\mathbf{k}}} \right] \nonumber \\
&= \prod_{\mathbf{k}} \left[ \sum_{n_\mathbf{k}=0}^{\infty} \delta_{n_\mathbf{k}} e^{-\beta (n_\mathbf{k}+1/2) E_\mathbf{k}} \right] \nonumber
\end{align}
where the integral in $\varphi_\mathbf{k}$ is solved using a change of variable and the residue theorem, as explained in the previous section. The final result is the same as the one that can be obtained using the base of number eigenstates for the evaluation of the trace. This calculation is intended to show the possibility of using the counting operator in the partition function in a context of quantum field theory.

\section{Ewkons and dark energy}

We can consider a quantum field with statistical weights
\begin{equation}
\delta_{n_\mathbf{k}} = \left\{ \begin{array}{cl}
0 & \text{if } n_\mathbf{k}<\sigma \\ 
1/(n_\mathbf{k}-\sigma)! & \text{if } n_\mathbf{k}\ge \sigma
\end{array}    \right.,
\end{equation}
associated to the number state with mode $\mathbf{k}$, such that ewkon statistics is obtained: $\mathcal{Z}_\text{ewk} = \prod_{\mathbf{k}} \mathcal{Z}_\mathbf{k}$ with
\begin{equation}
\mathcal{Z}_\mathbf{k} = e^{-\beta E_\mathbf{k} \sigma}\,\exp(e^{-\beta E_\mathbf{k}}).
\end{equation}
The vacuum energy factor, $e^{-\beta E_\mathbf{k}/2}$, was not included. It leads to inconsistencies in the evaluation, at a cosmological level, of, for example, the average energy of fermions or bosons, and is accordingly removed; see, e.g., \cite[p.\ 19]{kapusta}. Here, the same prescription is used for ewkons. Nevertheless, in the case of ewkons the removal of the vacuum energy factor does not qualitatively change thermodynamic properties; its inclusion is equivalent to make the replacement $\sigma \rightarrow \sigma +1/2$ in the following results. 

In order to evaluate the average energy density and pressure we need
\begin{align}
\frac{1}{V} \ln \mathcal{Z}_\text{ewk} &= \frac{1}{V} \sum_{\mathbf{k}} \ln \mathcal{Z}_\mathbf{k} \nonumber \\
&= \frac{1}{(2\pi)^3} \int d\mathbf{k}\; \ln \mathcal{Z}_\mathbf{k} \nonumber \\
&= \frac{1}{2\pi^2} \int_0^{k_\text{max}} dk \; k^2\, (e^{-\beta E_\mathbf{k}} - \beta E_\mathbf{k} \sigma)
\end{align}
where in the large volume limit the sum in $\mathbf{k}$ goes over to an integral, and an upper limit $k_\text{max}$ was included for the absolute value of $\mathbf{k}$ to avoid divergences; it is equivalent to an ultraviolet cutoff. 

Let us consider the most simple situation: a massless field, $m=0$, with $\mu=0$. 
The energy density is
\begin{equation}
\rho = - \frac{1}{V} \frac{\partial \mathcal{Z}_\text{ewk}}{\partial \beta}
\end{equation}
and the pressure is
\begin{equation}
P = \frac{1}{\beta V} \ln \mathcal{Z}_\text{ewk}.
\end{equation}
The upper limit for the energy is $E_\text{max} = k_\text{max}$. Assuming that $\beta E_\text{max} \gg 1$, we get
\begin{align}
\rho &\simeq \frac{E_\text{max}^4 \sigma}{8\pi^2}\left(1 + \frac{24}{\beta^4 E_\text{max}^4\sigma} \right) \label{rho} \\
P &\simeq -\frac{E_\text{max}^4 \sigma}{8\pi^2}\left(1 - \frac{8}{\beta^4 E_\text{max}^4\sigma}\right).
\end{align}
The parameter $w_\text{ewk}$, that represents the cosmological equation of state for ewkons, is
\begin{equation}
w_\text{ewk} = \frac{P}{\rho} \simeq -1 + \frac{32}{\beta^4 E_\text{max}^4\sigma}.
\label{wewk}
\end{equation}
The accelerated expansion of the universe implies a negative value of $w$, mainly due to the presence of dark energy. According to Table 3 in Ref.\ \cite{planck}, it is smaller than $-0.94$. Assuming that dark energy has the statistics of ewkons, and knowing that its energy density is $4\;10^9$ eV/m$^3$ \cite{beringer}, using \eqref{rho} we can obtain $E_\text{max} \sigma^{1/4} \simeq 0.028$ eV, a rather small value compared to the mass of elementary particles, but two orders of magnitude larger than the present value of $1/\beta$ ($1/\beta \simeq 2.4\;10^{-4}$ eV); and, using \eqref{wewk}, we have $w_\text{ewk} \simeq - 0.9999998$. 


\section{Conclusions}
\label{conclusions}

The inclusion of the counting operator $\hat{\delta}$ in the definition of the partition function allows to go beyond Fermi-Dirac or Bose-Einstein statistics. It is diagonal in the base of number states. An eigenvalue equal to 0 indicates that the corresponding number state is not allowed, and a value different from 0 represents its statistical weight.

Using this modified definition of the partition function, particles without spin, described through a free scalar field $\phi$, may obey to statistics other than Bose-Einstein's. Assuming that non-interacting particles have free diffusion in energy space, it has been shown that the possible statistics are the corresponding to bosons, fermions, classical particles and ewkons \cite{hoyusist}. The statistics of ewkons turns out to be particularly interesting for the description of dark energy, since a gas of ewkons has negative pressure, and a negative value of parameter $w$ that is necessary for an understanding of the accelerated expansion of the universe. This possibility was analyzed in the context of quantum field theory.

There is no contradiction with the spin-statistics theorem. According to this theorem, special relativity restricts the possible creation and annihilation operators to those that have commutation or anti-commutation relations; see, e.g., \cite[ch.\ 4]{srednicki}. The usual definition \eqref{defZ} of the partition function implies that we only have Bose-Einstein or Fermi-Dirac statistics for commutation or anti-commutation relations respectively. Nevertheless, the definition of the partition function with the counting operator \eqref{qpartfun} leads to other possible statistics for creation and annihilation operators, associated to a scalar field, that satisfy the commutation relation (see the appendix). The choice of the definition depends on its utility to describe nature. Here it is argued that generalized statistics of particles without spin may be useful to describe dark energy.

\section{Appendix}

The number of particles operator for mode $\mathbf{k}$ is $\hat{n}_\mathbf{k} = \hat{a}^\dagger_\mathbf{k} \hat{a}_\mathbf{k}$, with the annihilation operator given by
\begin{equation}
\hat{a}_\mathbf{k} = \frac{1}{\sqrt{2E_\mathbf{k} V}} (E_\mathbf{k} \hat{\phi}_\mathbf{k} + i \hat{\pi}_\mathbf{k}).
\end{equation}
From the canonical commutation relation $[\hat{\phi}(\mathbf{x}),\hat{\pi}(\mathbf{x}')]=i \delta^3(\mathbf{x}-\mathbf{x}')$ it can be shown that 
\begin{equation}
[\hat{\phi}^\dagger_\mathbf{k}, \hat{\pi}_{\mathbf{k}'}] = i V \delta_{\mathbf{k},\mathbf{k}'},
\end{equation}
where $\delta_{\mathbf{k},\mathbf{k}'}$ is the Kronecker delta and $V$ is the system's volume. This commutation relation implies that
\begin{equation}
[\hat{a}_{\mathbf{k}},\hat{a}^\dagger_{\mathbf{k}'}] = \delta_{\mathbf{k},\mathbf{k}'}.
\end{equation}
Let us notice that this commutation relation holds not only for Bose-Einstein statistics, but to any statistics obtained with a counting operator $\hat{\delta}_\mathbf{k}$. Using these relations we can obtain
\begin{align}
\hat{H}_\mathbf{k} &= (\hat{a}^\dagger_\mathbf{k} \hat{a}_\mathbf{k} +1/2)E_\mathbf{k} \nonumber \\
&= \frac{1}{2V}\left( \hat{\pi}_\mathbf{k}\hat{\pi}_\mathbf{k}^\dagger + E_\mathbf{k}^2\, \hat{\phi}_\mathbf{k}\hat{\phi}_\mathbf{k}^\dagger \right) + \frac{iE_\mathbf{k}}{2V}(\hat{\phi}_\mathbf{k}^\dagger \hat{\pi}_\mathbf{k} - \hat{\phi}_{-\mathbf{k}}^\dagger \hat{\pi}_{-\mathbf{k}}).
\end{align}
The term $\hat{\phi}_\mathbf{k}^\dagger \hat{\pi}_\mathbf{k} - \hat{\phi}_{-\mathbf{k}}^\dagger \hat{\pi}_{-\mathbf{k}}$ can be ignored in the calculations since it vanishes when the sum on $\mathbf{k}$ is performed.

\section*{Acknowledgments}
I acknowledge useful discussions with Pablo Sisterna and Ariel Megevand. This work was partially supported by Consejo Nacional de Investigaciones Cientí\-ficas y Técnicas (CONICET, Argentina, PIP 0021 2015-2017).


\begin{thebibliography}{00}
	
\bibitem{kapusta} J. I. Kapusta and C. Gale, \textit{Finite-Temperature Field Theory}, Cambridge University Press, 2006.

\bibitem{laine} M. Laine and A. Vuorinen, \textit{Basics of Thermal Field Theory}, Springer, 2016.

\bibitem{hogan} J. Hogan, Welcome to the Dark Side, Nature \textbf{448} (2007) 240-245 .

\bibitem{planck} P.A.R. Ade, \textit{et al.}, Planck 2015 results. XIV. Dark energy and modified gravity, Astron. Astrophys. 594 (2016) A14. 

\bibitem{amendola} L. Amendola, Coupled quintessence, Phys. Rev. D \textbf{62} (2000) 043511.

\bibitem{gott} Zachary Slepian, J. Richard Gott III, Joel Zinn, A one-parameter formula for testing slow-roll dark energy: observational prospects, MNRAS \textbf{438} (2014) 1948-1970.

\bibitem{hoyusist} M. Hoyuelos and P. Sisterna, Quantum statistics of classical particles derived from the condition of a free diffusion coefficient, Phys. Rev. E \textbf{94} (2016) 062115.

\bibitem{hoyuelos} M. Hoyuelos, From creation and annihilation operators to statistics, Physica A \textbf{490} (2018) 944.

\bibitem{gentile} G. Gentile, Osservazioni sopra le statistiche intermedie, Nuovo Cimento \textbf{17} (1940) 493.

\bibitem{green} H. S. Green, A Generalized Method of Field Quantization, Phys. Rev. \textbf{90} (1953) 270.

\bibitem{katsura} Shigetoshi Katsura, Katsuji Kaminishi, and Sakari Inawashiro, Intermediate Statistics, Journal of Mathematical Physics \textbf{11} (1970) 2691.

\bibitem{wilczek} F. Wilczek, Quantum Mechanics of Fractional-Spin Particles, Phys. Rev. Lett. \textbf{49} (1982) 957.

\bibitem{greenberg} O. W. Greenberg, Example of Infinite Statistics, Phys. Rev. Lett. \textbf{64} (1990) 705.


\bibitem{haldane}  F. D. M. Haldane, ``Fractional statistics'' in arbitrary dimensions: A generalization of the Pauli principle, Phys. Rev. Lett. \textbf{67} (1991) 937.

\bibitem{isakov} S. B. Isakov, Generalization of Quantum Statistics
in Statistical Mechanics, International Journal of Theoretical Physics \textbf{32} (1993) 737.

%

\bibitem{isakov2} S. B. Isakov, D. P. Arovas, J. Myrheim, and A. P. Polychronakos, Thermodynamics for fractional exclusion statistics, Physics Letters A \textbf{212} (1996) 299.

\bibitem{poly} A. P. Polychronakos, Probabilities and path-integral realization of exclusion statistics, Physics Letters B \textbf{365} (1996) 202.

%

\bibitem{anghel} D. V. Anghel, Fractional exclusion statistics in general systems of interacting particles, Physics Letters A \textbf{372} (2008) 5745.


\bibitem{dai} Wu-Sheng Dai and Mi Xie, Calculating statistical distributions from operator relations: The statistical distributions of various intermediate statistics, Annals of Physics \textbf{332} (2013) 166.



\bibitem{khare} Avinash Khare, \textit{Fractional Statistics and Quantum Theory}, World Scientific Publishing, 2005.
%

\bibitem{beringer} J. Beringer \textit{et al.} (Particle Data Group), Review of particle physics, Phys. Rev. D 86 (2012) 010001.

\bibitem{srednicki} M. Srednicki, \textit{Quantum field theory}, Cambridge University Press, 2010.





%
%
%
%
%
%
%
%
%
%
%
%
%

\end{thebibliography}
\end{document}